\documentclass[prb, twocolumn]{revtex4}
\usepackage{amsmath, amsfonts, amsthm, bbm,bm}
\usepackage{natbib} 
\usepackage{graphicx}
\newcommand{\Exe}{{E\!\otimes\! e}}         
\renewcommand{\[}{\begin{equation}}
\renewcommand{\]}{\end{equation}}
\newcommand{\ITPFUB}{Institut f\"ur Theoretische Physik, Freie Universit\"at Berlin, 14195 Berlin, Germany}
\def\fbkt#1#2#3{\mathinner{\langle{#1}\,\lvert\,#2\,\lvert\,#3\rangle}} 
\def\ket#1{\mathinner{\lvert{#1}\rangle}}
\newcommand{\mb}[1]{\text{\boldmath ${#1}$}}
\newcommand{\abs}[1]{\lvert#1\rvert}

\begin{document}
\title{Berry-Phase Effects in Transport through Single Jahn-Teller Molecules}
\author{Maximilian G.\ Schultz}\email{mschultz@physik.fu-berlin.de}
\author{Tamara S.\ Nunner}
\author{Felix von Oppen}
\affiliation{\ITPFUB}
\date{\today}
\begin{abstract} 

  The vibrational modes of Jahn-Teller molecules are affected by a Berry phase that is associated
  with a conical intersection of the adiabatic potentials. We investigate theoretically the effect
  of this Berry phase on the electronic transport properties of a single $\Exe$ Jahn-Teller molecule
  when the tunneling electrons continually switch the molecule between a symmetric and a Jahn-Teller
  distorted charge state. We find that the Berry phase, in conjunction with a spectral trapping
  mechanism, leads to a current-blockade even in regions outside the Coulomb blockade. The blockade
  is strongly asymmetric in the gate voltage and induces pronounced negative differential
  conductance. 

\end{abstract} 
\maketitle

\section{Introduction} 

The vision of molecular electronics has stimulated great interest in both the
experimental~\cite{Park00,Ruitenbeek_H2,Ralph,Zant,Yacoby,Natelson} and the
theoretical~\cite{Gorelik98,Braig03,Mitra04,Kaat04,Nowack05,Koch05,Koch06,Donarini06} understanding
of electronic transport through single molecule devices. The coupling of electronic degrees of
freedom to few, well-defined molecular vibrations, a property that discriminates transport through
single molecules from that through other nanostructures such as quantum dots, sets the stage for
observing novel quantum-transport phenomena.~\cite{Nitzan} One of the simplest and most intensively
studied models of such a device is a molecule with a single electronic level and a one-dimensional
potential energy surface for the nuclear displacements. This model already gives rise to a large
variety of interesting transport phenomena such as vibrational sidebands,~\cite{Braig03,Mitra04}
electron shuttles,~\cite{Gorelik98} Franck-Condon blockade,~\cite{Koch05} avalanche-like
transport,~\cite{Koch05} pair tunneling,~\cite{Koch06} and dynamical symmetry
breaking.~\cite{Donarini06}  Generalizations of this model to two or more electronic levels have
been suggested in order to account for non-degenerate but competing molecular states~\cite{Nowack05}
and for degenerate electronic states on a mole\-cular dimer.~\cite{Kaat04}

Extensions of the simple molecular model to higher-dimensional potential energy surfaces may seem to
be straightforward without yielding qualitatively new physics. This is, however, \textit{not} the
case for molecules with symmetry-induced degeneracies of both electronic and vibrational states. For
such molecules, the Jahn-Teller effect\cite{Bersuker} leads to nontrivial vibrational dynamics,
originating in the conical intersections of adiabatic potentials and their associated geometric
(Berry) phases. We expect the transport properties to be particularly sensitive to this intriguing
dynamics when the tunneling of electrons between molecule and leads switches the molecule between a
symmetric and a Jahn-Teller distorted charge state.

In this paper, we investigate the consequences of such a Berry phase for transport through single
Jahn-Teller molecules. We focus on the $\Exe$ Jahn-Teller effect, for it is the simplest example
with a conical intersection of the potential energy surfaces. This type of Jahn-Teller effect occurs
for a large variety of triangular (X$_3$), tetrahedral (ML$_4$---a transition metal ion M surrounded
by a tetrahedron of ligands L), and octahedral molecules or complexes ML$_6$. We show in section
\ref{ssec:SelectionRule} that the Berry phase induces a nontrivial selection rule for transitions
between vibronic excitations of different molecular charge states.  This selection rule, in combination with generic properties of the
energy spectrum of the Jahn-Teller molecule, leads to the formation of trapping states in the
sequential tunneling limit (Section \ref{sec:Theory}).  The consequences are a current-blockade
outside the Coulomb blockade window, asymmetry of the current-voltage characteristics, and strong
negative differential conductance (Section \ref{sec:Rateequations}). The appearance of trapping states also strengthens the role
of higher order processes, which introduce a second time scale that is much larger than the scale on
which sequential tunneling takes place. The interplay of these two scales renders the details of the
tunneling dynamics irrelevant and has a non-negligible effect on the line shape of the differential
conductance peaks. We conclude the paper with a discussion of some experimental issues (Section
\ref{sec:Perturbations}) and an outlook to interesting problems related to electronic transport
through Jahn-Teller molecules (Section \ref{sec:Conclusion}).

\begin{figure}
\begin{minipage}{.35\linewidth}
    \includegraphics[width=\linewidth]{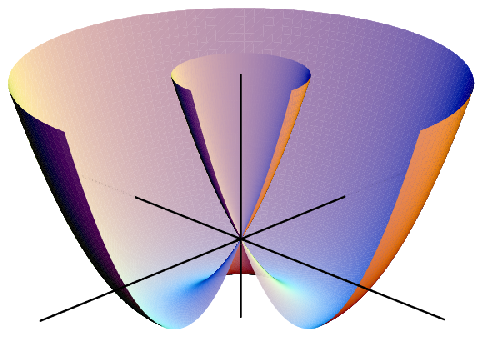}
 \end{minipage}
\begin{minipage}[c]{.35\linewidth}
    \includegraphics[width=\linewidth]{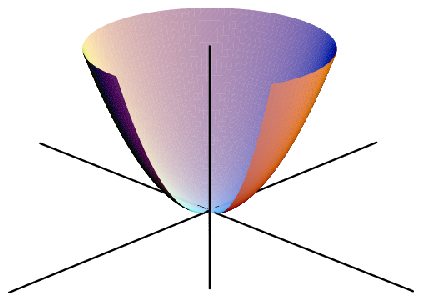}
 \end{minipage}\\
  \begin{minipage}[c]{.35\linewidth}
    \includegraphics[height=1.1\linewidth]{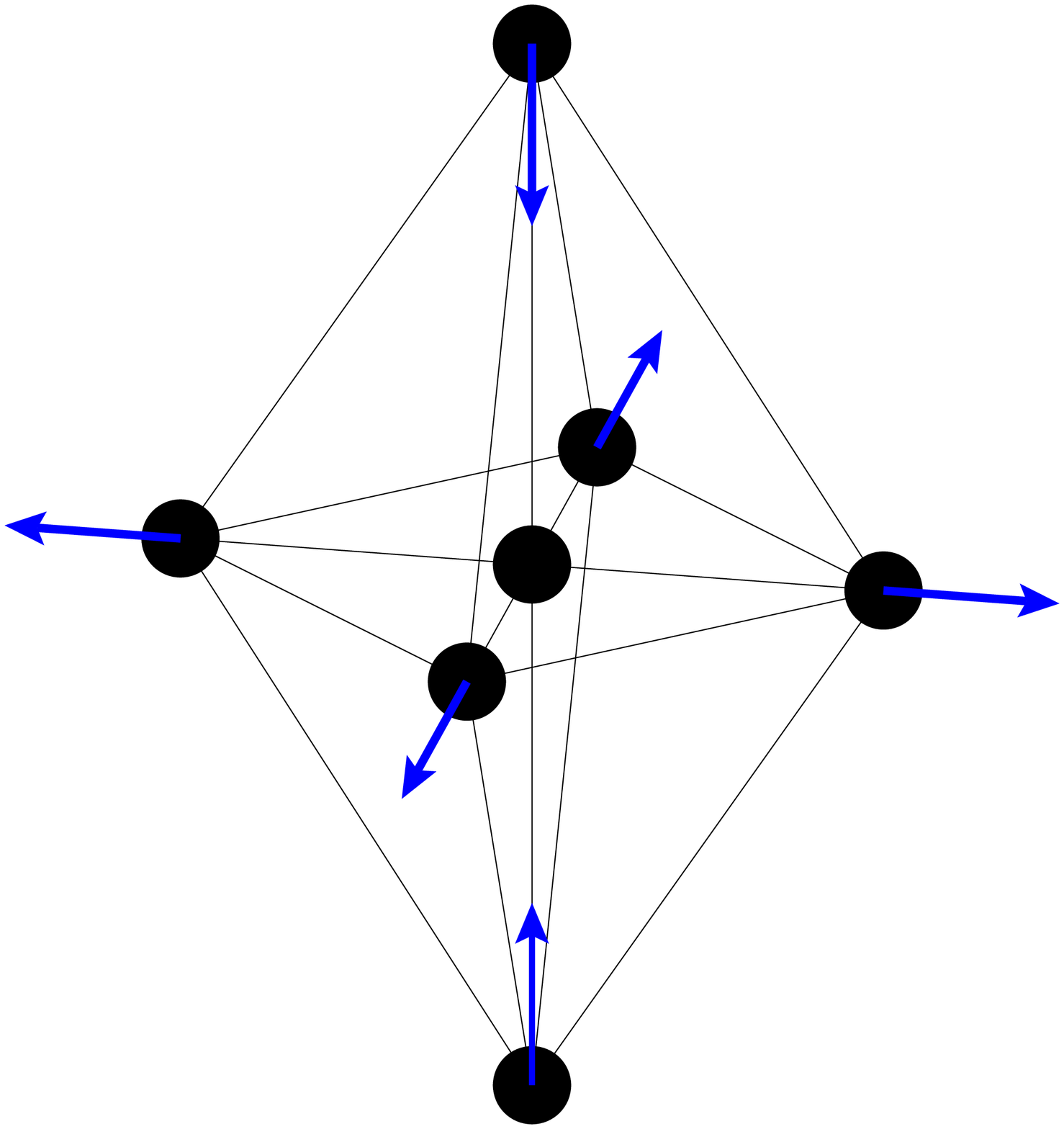}
  \end{minipage}
  \begin{minipage}[c]{.35\linewidth}
    \includegraphics[height=1.1\linewidth]{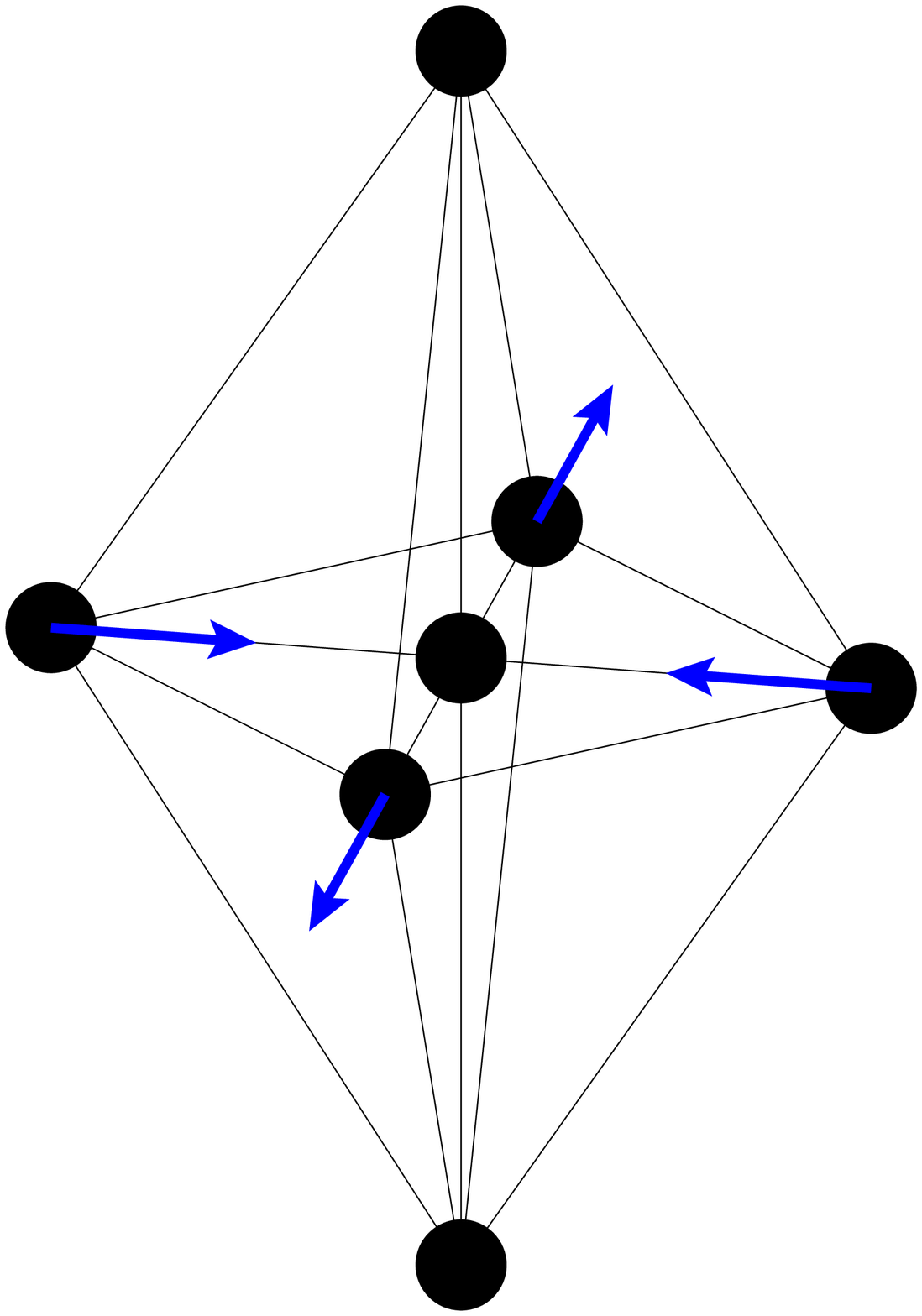}
  \end{minipage}
    \caption{{\bf Top:} Adiabatic potentials of the $\Exe$ Jahn-Teller
    distorted (charged) molecule (left) and the undistorted (neutral) molecule (right) as
    a function of $(Q_\theta, Q\varepsilon)$. {\bf Bottom:} Displacement
    patterns of the degenerate $\theta$ and $\varepsilon$ vibrations
    for octahedral ML$_6$ molecules.
\label{fig:mexican_hat}}
\end{figure}

\section{Jahn-Teller molecule}\label{sec:JTE}
\subsection{Model}

The $\Exe$ Jahn-Teller effect generally occurs in molecules where two degenerate electronic states
are coupled to two degenerate vibrational modes $\theta$ and $\varepsilon$. In the case of an
octahedral ML$_6$ molecule, the relevant electronic states are of $d_{x^2-y^2}$- and
$d_{3z^2-r^2}$-type, and the degenerate vibrations correspond to the symmetric oscillations of the
ligands as they are pictured in Fig~\ref{fig:mexican_hat}. For definiteness, we assume that without
an excess charge on the molecule the degenerate electronic $E$ states are empty.~\footnote{Our
results are independent of this choice as long as the tunneling electron switches the molecule
between its Jahn-Teller distorted and symmetric states.} With this choice, the molecular Hamiltonian
is~\cite{Bersuker} 
\[\label{eq:H_exe} 
\begin{split} H_{E\otimes e}
  = & eV_\text{g} n_\text{d}  + \frac{\hbar\omega}{2}(P_\theta^2 + Q_\theta^2 +
  P_\varepsilon^2 + Q_\varepsilon^2)\\&+ \lambda \hbar\omega\left(Q_\theta
\sigma_x + Q_\varepsilon \sigma_y\right)n_\text{d}. 
\end{split} 
\] 
We use a pseudospin notation with Pauli matrices $\sigma_i$ for the electronic states $\ket{+}$,
$\ket{-}$. The dimensionless strength of the linear coupling of electronic and vibrational degrees
of freedom is $\lambda$, and $\omega$ is the frequency of the vibrational mode with effective
mass $M$ and oscillator length $\ell_\text{osc} = (\hbar/M\omega)^{1/2}$. The dimensionless
coordinates $Q$ and momenta $P$ of the normal modes $\theta$ and $\epsilon$ are measured in units of
$\ell_\text{osc}$. The gate voltage $V_\text{g}$ tunes the energy of the degenerate electronic
levels. $V_\text{g}=0$ means that for vanishing bias, the molecular levels are degenerate with the
Fermi levels of the electrodes. Electronic occupations different from $n_\text{d}=0$ (neutral molecule) and $n_\text{d}=1$
(charged molecule) are effectively projected out by assuming a large charging energy $U$.

\begin{figure}
\begin{minipage}{.9\linewidth}
    \includegraphics[width=\linewidth]{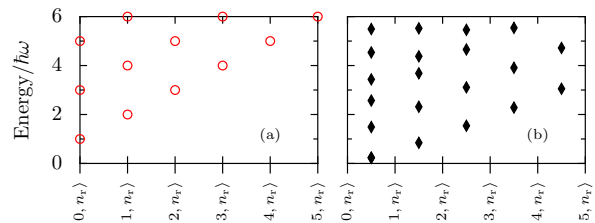}
 \end{minipage}
 \caption{Vibrational spectra (a) of the neutral and (b) of the charged molecule vs. Pseudo angular
 momentum. Only energies for positive pseudo angular momenta are shown as the spectra are symmetric under
 the reflection $l\mapsto -l$.\label{fig:spectra_a}}
\end{figure}

The potential surfaces of both charge states are $U(1)$ symmetric in the $(Q_\theta, Q_\varepsilon)$
plane and thus have a conserved angular momentum quantum number. We shall refer to this angular
momentum as a {\em pseudo} angular momentum to make clear that it does \textit{not} describe
rotations in real space but rather a continuous change of the shape of the molecule's distortion. A
vibronic state of the neutral molecule is labeled by its \textit{integer} pseudo angular momentum $l$ and
radial excitation $n_\text{r}$. As it is well-known from the occupation number
representation of the two-dimensional harmonic oscillator, its energy depends on the total number of
vibrational quanta no matter in which direction of the oscillator they are excited. The energy
manifold $E=(N+1)\hbar\omega$ is $N$-fold degenerate. In the pseudo angular momentum representation,
the degeneracy is reflected in the relation $E_{l, n_\text{r}} = E_{l+2, n_\text{r}-1}$, compare Fig.~\ref{fig:spectra_a}~(a).

In the charged configuration, on the contrary, with an explicit coupling of vibrations to the
pseudospin, it is the \textit{total} pseudo angular momentum $\mb{j}=\mb{l}+\frac{1}{2}\mb{\sigma}$
that is conserved. That the conserved pseudo angular momentum must be a \textit{half-integer}
quantity can also be seen by inspection of the adiabatic potentials, which are computed by
diagonalizing the potential energy of $H_{E\otimes e}$ for fixed nuclear displacements $(Q_\theta,
Q_\varepsilon)$.  The two sheets of the $n_\text{d}=1$ adiabatic potential have, in contrast to the
single sheet of the neutral state, a conical intersection at the origin. Along any loop enclosing
the conical intersection, the electronic pseudospin of the adiabatic eigenstates rotates by
$2\pi$.~\cite{Berry84} The adiabatic eigenstates consequently acquire a Berry phase of $\pi$ along
such closed loops, which must be offset by the vibrational wavefunction in order to make the full,
electronic plus vibrational, wavefunction single valued. This implies {\em anti}\!\:periodic
boundary conditions for the vibrational wavefunction and thus \textit{half-integer} pseudo angular
momentum $j$. The spectrum of the charged configuration is plotted in Fig.~\ref{fig:spectra_a}~(b)
for an intermediate value of the electron-phonon coupling, $\lambda=1$. In contrast to the
regularity of the oscillator spectrum, the Jahn-Teller effect lifts the large degeneracy; each state
is only doubly degenerate, $E_{j,n_\text{r}} = E_{-j,n_\text{r}}$. The Jahn-Teller effect also lowers the ground state
energies of each pseudo angular momentum column with respect to the harmonic oscillator energies,
which is one of the keys to the current-blockade mechanism discussed in this paper.

The molecular system is coupled to two non-interacting Fermi-gas electrodes labeled by $\alpha =
\text{L}, \text{R}$. We define electron annihilation operators $c_{\alpha\bm{k}s}$ for a particle in
lead $\alpha$ with wavevector $\bm{k}$ and spin $s$, and $d_{\pm, s}$ for a particle on the
molecule. The tunneling between the molecule and the electrodes is described by a conventional
tunnel Hamiltonian, $H_\text{T} =
t_0\smash{\sum_{\alpha,\mb{k},s}}c_{\alpha\mb{k}s}(\smash{d^\dagger_{+,s}} +
\smash{d^\dagger_{-,s})} + \text{h.c.}$ Jahn-Teller distortions usually occur on a sub-\AA
ngstr\"om scale,\cite{Bersuker} which is small compared with the scale on which electronic
wavefunctions fall off. For this reason, we can neglect the dependence of $t_0$ on the vibrational
coordinates. The full Hamiltonian of the single molecule device is 
\[
H = H_{E\otimes e} + H_{\text{leads}} + H_\text{T},
\]
where $H_\text{leads} = \sum_\alpha\sum_{\mb{k}s} \varepsilon_{\mb{k}}c_{\alpha\mb{k}s}^\dagger
c_{\alpha\mb{k}s}$ is the Hamiltonian of the electrodes. The states of the molecule will be labeled
$\ket{n_\text{d};j,n_\text{r}}$, $n_\text{d}$ is the charge on the molecule, $j$ is the pseudo angular
momentum ($l$ in the neutral state), and $n_\text{r}$ enumerates the radial excitations. Our discussion is restricted
to the weak tunneling case, where the tunneling induced broadening of
the molecular states $\Gamma$ is small compared with the temperature $T$, and the predominant
transport process is sequential tunneling of electrons. We will only consider a regime of intermediate
electron phonon coupling, where the system is not in the Franck-Condon blockade.\cite{Koch05} The
Franck-Condon blockade effect would not alter the validity of our conclusions but merely complicate
the discussion unnecessarily.

\subsection{Selection Rule}\label{ssec:SelectionRule}

Any computation of transport properties via an expansion of the system's reduced density matrix in
powers of the tunnel Hamiltonian requires the computation of the matrix
elements of $H_\text{T}$. An expansion of the Jahn-Teller states $\ket{1;j,n_\text{r}}$ in terms of harmonic
oscillator eigenstates $\ket{l, n_\text{r}'}$ times the two-dimensional electronic manifold $\ket{\pm}$
shows that the electron-phonon interaction due to the $\Exe$ Jahn-Teller effect only couples the states $\ket{+}\!\ket{l,n_\text{r}'}$ with
the states $\ket{-}\!\ket{l+1,n_\text{r}''}$ of the non-interacting system.\cite{Longuet} The associated half-integer pseudo angular momentum number of the
Jahn-Teller state is $j=l+\frac{1}{2}$. The eigenstates of the Jahn-Teller active charge state thus have
the general form
\[
\ket{1;j,n_\text{r}} =
\sum_{\tilde{n}_\text{r}}\left[A^{jn_\text{r}}_{\tilde{n}_\text{r}}\ket{+}\!\ket{j-\frac{1}{2},
\tilde{n}_\text{r}} +
B^{jn_\text{r}}_{\tilde{n}_\text{r}}\ket{-}\!\ket{j+\frac{1}{2}, \tilde{n}_\text{r}} \right]
\]
for suitable expansion coefficients $A^{jn_\text{r}}_{\tilde{n}_\text{r}}$ and
$B^{jn_\text{r}}_{\tilde{n}_\text{r}}$. The matrix
elements $\fbkt{0;l,n_\text{r}'}{H_\text{T}}{1;j,n_\text{r}}$ will therefore vanish unless the pseudo angular momenta
fulfil
\[\label{eq:Selection_Rule}
j = l\pm\frac{1}{2}.
\]
This condition causes many zeroes in the tunnel matrix and has a great influence on the transport
properties of the device as will be shown in the following sections.

\section{Current blockade}\label{sec:Theory}

The main result of our paper is a current-blockade effect \textit{outside} the Coulomb blockade diamonds. In
certain areas of the $(V_\text{g}, V_\text{sd})$ plane, the system exhibits ``trapping states''
that, if populated, lock an excess charge on the molecule and, due to the large charging energy,
obstruct any stationary current to flow through the device. These states are vibrational eigenstates
of the Jahn-Teller active molecule with pseudo angular momentum $\vert j\,\vert>\frac{1}{2}$. The effect becomes
more pronounced for larger pseudo angular momenta. For certain voltage parameters that lie well
outside the Coulomb-blockaded region, an electron tunneling from the source electrode can excite the molecule
into such states, but it cannot tunnel out towards the drain electrode. The trapping effect itself
comes about, because these states become isolated in the tunneling dynamics: either a transition to
a neutral target state is forbidden by the selection rule or it is forbidden by conservation of
energy.

\begin{figure}[tb]
  \includegraphics[width=\linewidth]{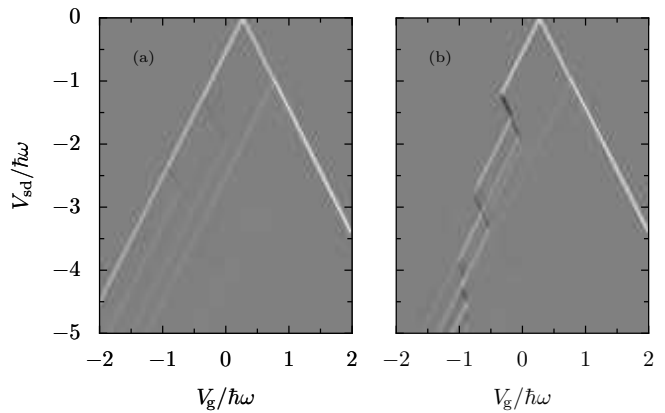}
  \caption{Differential conductance vs.\ gate voltage $V_\text{g}$ and source-drain
  voltage $V_\text{sd}$ for $\lambda = 1$, $k_\text{B}T = 0.01\,\hbar\omega$. (a) $\tau =
  10\,\tau_0$, (b) $\tau= 5000\,\tau_0$. White correspond to positive, Black to
  negative values. Neutral Gray is zero differential conductance.\label{fig:results}} 
\end{figure}

The implications of this effect for the differential conductance are the formation of structures
like those in Fig.~\ref{fig:results}: a suppression of the stationary current outside the Coulomb
blockaded region. The suppression, however, is not complete; a small stationary current is admitted
due to the presence of higher order processes such as co-tunneling, relaxation to thermal equilibrium, or mixing of
states by external perturbations, which effectively break the strict selection rule, allow to
release the system to a vibrational state that can then be left by tunneling. Since these processes
are of higher order in $t_0$, they define a second time scale $\tau$. This time scale, which
describes the time that the system will effectively spend in a trapping state, is much larger than the
time scale of sequential tunneling, given by $H_\text{T}$, $\tau_0^{-1} := 2\pi\nu t_0^2$, where
$\nu$ is the density of states in the electrodes.

\subsection{Formalism}

The technical and formal discussion of the trapping effect is subtle, because due to the high level
of degeneracy, one has to employ full Master equation dynamics instead of intuitive rate equations
for sequential tunneling. The qualitative picture can nonetheless be grasped by rate equations in
the pseudo angular momentum basis. 

We confine ourselves to the sequential tunneling regime, which means that transitions will only take
place between different charge states. We use an expansion of the reduced density matrix $\rho$ in
the tunneling Hamiltonian,\cite{Mitra04} employ both the Markov and the Born approximation, and
eventually arrive at a Liouville-von Neumann equation, $\dot{\rho} = -i[H_\text{leads}+H_{E\otimes
e}, \rho] + \mathcal{L} \rho$. The operator $\mathcal{L}$ has the well-known Lindblad form for
dissipative linear operators and describes the tunneling in second order perturbation
theory.\cite{Blum81} Since we are only interested in the stationary transport properties, we can safely
neglect those off-diagonal matrix elements of $\rho$ that belong to states of different energies and
thus effectively produce a rate equation dynamics between different energy manifolds. The stationary
von Neuman equation reduces with this approximation to $\mathcal{L}\rho = 0$. But since the system
under consideration has degenerate states, the equations can, at this point, \textit{not} be further
simplified to rate equations for the states $\ket{1;j,n_\text{r}}$ and $\ket{0;l,n_\text{r}'}$. We do, however, know
that the density matrix is a self-adjoint operator and hence admits an eigenbasis in the
degenerate subspaces for \textit{each} set of voltage parameters $(V_\text{sd}, V_\text{g})$. In this basis,
the von Neumann equation assumes the form of a rate equation, where the rate for a transition
between two eigenstates $\ket{\beta}$ and $\ket{\gamma}$ is given by a golden-rule expression,
\[\label{eq:GoldenRule}
M^{\beta \mapsto \gamma} := \frac{2\pi}{\hbar}\nu t_0^2 \abs{{\fbkt{\gamma}{d^\dagger_+
+d^\dagger_-}{\beta}}}^2,
\]
which is of order $\mathcal{O}(H_\text{T}^2)$, multiplied by a Fermi function that accounts for the
availability of an electron or a hole in the leads. If we want to compute the rates between different energy manifolds
in the tunneling problem, we shall consider the, unknown, diagonal basis of the degenerate subspaces
of $\rho$ at fixed voltages. Let $\ket{\gamma}=\sum a_{n_\text{r},l}\ket{l,n_\text{r}}$ be one of these states in the neutral configuration 
and $\ket{\beta} = a_j \ket{j,n_\text{r}'} + a_{-j}\ket{-j,n_\text{r}'}$ one in the charged configuration. The Franck-Condon factor is
\[
\begin{split}
 \abs{\fbkt{\gamma}{H_\text{T}}{\beta}}^2 & = \abs{\sum_{n_\text{r}l, \pm j}a^\ast_{n_\text{r},l}b_j\fbkt{n_\text{r},l}{H_\text{T}}{n_\text{r}',j}}^2\\
  & = \abs{\sum_{\pm j} a^\ast_{n_\text{r},j\pm\frac{1}{2}}b_j\fbkt{n_\text{r},j\pm\frac{1}{2}}{H_\text{T}}{n_\text{r}',j}}^2.
\end{split}
\]
The selection rule, which has been shown to be strict in the pseudo angular momentum basis, has 
thus implications on the transition matrix elements whatever the eigenbasis of $\rho$ is. If the angular momentum of a charged state
$\ket{\beta}$ is too large compared with those present in a neutral state $\ket{\gamma}$, a transition may not be
allowed. We will show in the next section that such forbidden transitions between energy
manifolds are the cause of the current-blockade that we observe analytically and numerically.

\subsection{Trapping states}

\begin{figure}
\begin{minipage}{.9\linewidth}
    \includegraphics[width=\linewidth]{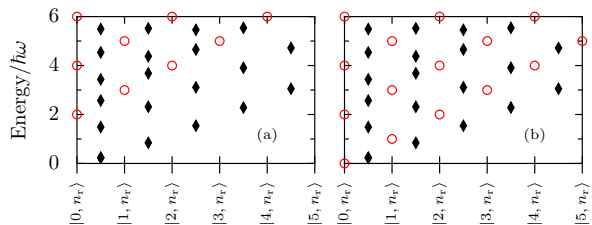}
\end{minipage}

 \caption{Angular-momentum resolved spectra of the Jahn-Teller molecule ($\lambda = 1$) (diamonds)
 and the neutral molecule  (circles). Radial excitations are grouped for each pseudo angular
 momentum.  Panel~(a) shows the situation for tunneling at the source electrode, panel~(b) the
 situation at the drain electrode. The chosen parameters are $eV_\text{sd} = 2\hbar\omega$,
 $eV_\text{g} = -0.5\hbar\omega$. Consider panel~(a). An electron can only tunnel from the source
 electrode and induce the vibronic transition $\ket{0;q}\to\ket{1;q'}$ if the difference in the
 electronic energy, $\frac{1}{2}eV_\text{sd}-eV_\text{g}$, is larger than the difference in the
 vibronic energy, $E_{1q'}-E_{0q}$. More precisely, the Fermi function $\smash{f\bigl(e(V_\text{g} -
 \frac{1}{2}V_\text{sd}) + E_{1q'}-E_{0q}\bigr)}$ has to be positive. By shifting the neutral
 spectrum by $\frac{1}{2}eV_\text{sd}$ and the charged spectrum by $eV_\text{g}$, this relation
 translates into the intuitive rule that in the diagram the target state of a tunnel process has to
 lie below the initial state. Energetically allowed transitions  are therefore directed downwards in
 the figure. A horizontal transition corresponds to the borderline case of an electron sitting
 exactly at the Fermi level of the electrode. That conservation of energy has the form of an
 inequality is due to the filled continuum of states in the Fermi-gas electrode.  
 Panel~(b) shows the corresponding energetic situation for sequential tunneling processes at the
 drain electrode.  The presence of trapping states can be seen clearly.  \label{fig:spectra_b}}

\end{figure}

\begin{figure}
\begin{minipage}{.9\linewidth}
    \includegraphics[width=.75\linewidth]{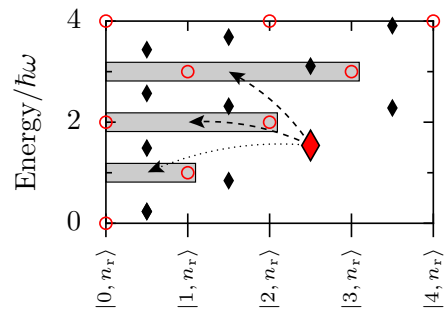}
\end{minipage}

 \caption{Angular momentum resolved spectra for tunneling to the drain electrode, $eV_\text{sd} =
 2\hbar\omega$, $eV_\text{g} = -0.5\hbar\omega$. The trapping state $\ket{\beta} =
 \ket{1;\frac{5}{2},0}$ is marked in red. The three nearest energy manifolds of the uncharged system are shaded in gray. The dashed
 arrows show transitions that are forbidden, because the target manifold's energy is too large. The
 dotted arrow shows a transition that is forbidden due to the selection rule.\label{fig:spectra_d}}
\end{figure}

Considering a sufficiently large bias and a gate voltage that places the ground state of the charged
molecule slightly above the Fermi level of the drain electrode, the Fermi functions for relevant
excitations are $1-f_\text{source} = f_\text{drain} \approx 0$, and we only need to discuss the
rates for an electron to leave the molecule towards the drain electrode. In this configuration, the
system is not in the Coulomb blockaded region and ought to be conducting. This is, however,
\textit{not} the case due to the selection rule Eq.~(\ref{eq:Selection_Rule}) and the spectral
structure of the molecule.  Consider Fig.~\ref{fig:spectra_b}, in which the vibronic spectra of both
charge states have been overlaid and shifted  by $eV_\text{g}$ and
$\pm\frac{1}{2}eV_\text{sd}$ along the energy axis to account for the electronic
energies of the charged and the neutral state respectively. Conservation of energy then implies that
transitions are energetically possible if the charged target state lies below the neutral initial state
in Fig.~\ref{fig:spectra_b}~(a) and vice versa in Fig.~\ref{fig:spectra_b}~(b), see the figure caption for
more details. 
Take for example the state $\ket{\beta} := \ket{1;\frac{5}{2},0}$, the charged state with
pseudo angular momentum $j=\frac{5}{2}$ and radial ground state. Once the system occupies this
state, it will stay there forever; the electron cannot tunnel to the drain electrode, see
Figs.~\ref{fig:spectra_b}~and~\ref{fig:spectra_d}. Conservation of energy requires that any target
state has to have less energy than $\ket{\beta}$, so only $l\le 1$ and suitable $n$ are possible.
The selection rule Eq.~(\ref{eq:Selection_Rule}) only allows the dynamics to connect neighboring
columns of Fig.~\ref{fig:spectra_b}. But then all transition rates are zero. The only states that
have non-vanishing Franck-Condon factors belong to energy manifolds that due to the specific choice
of voltages are out of reach energetically. The state $\ket{\beta}$ is called absorbing or ``trapping state'',
that is to say there is no escape rate but yet a finite rate for populating the state. The latter of
the two properties of $\ket{\beta}$ is easily verified with Fig.~\ref{fig:spectra_b} (a), which
shows the shift of the two spectra for tunneling processes from the source electrode. In this
configuration there are enough states from which $\ket{\beta}$ can be populated, for instance the
$E=4\hbar\omega$ manifold.  

\begin{figure}[b]
\begin{minipage}{.9\linewidth}
    \includegraphics[width=\linewidth]{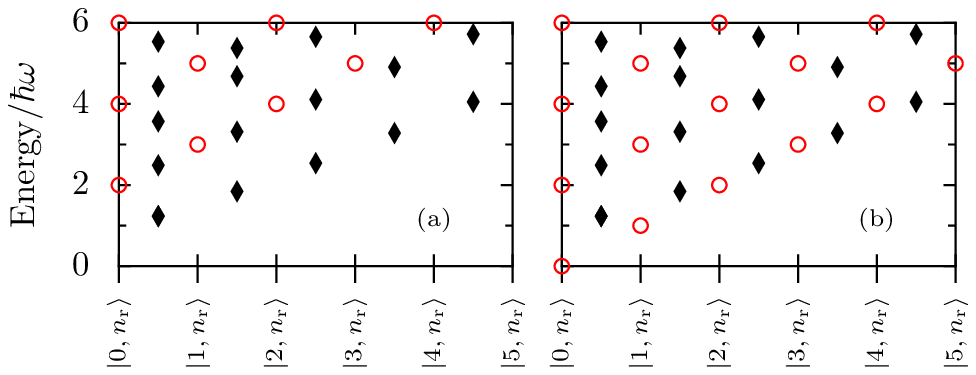}
\end{minipage}

\caption{Angular-momentum resolved spectra like in Fig.~\ref{fig:spectra_b}. In this case the
Jahn-Teller molecule is placed slightly below the chemical potential of the source electrode. The
chosen parameters are $eV_\text{sd} = 2\hbar\omega$, $eV_\text{g} = 0.5\hbar\omega$. Contrary to the
situation shown in Fig.~\ref{fig:spectra_b}, there are no trapping states.\label{fig:spectra_c}}

\end{figure}

The energy threshold above which a trapping state can be reached from the vibronic ground-state of
the neutral system marks the onset of the current-blockade in the $dI/dV$ diagram. It can thus cause
negative differential conductance peaks, like those in Fig.~\ref{fig:results}, if the
system were conducting below the threshold. Positive differential conductance peaks, on the
contrary, can be observed for the energy threshold above which the trapping state acquires a finite
escape rate and is no longer trapping.  The current-blockade effect, as it has been discussed in the
previous paragraph also produces a significant asymmetry of the $dI/dV$ diagram with respect to
$V_\text{g}$. In Fig.~\ref{fig:spectra_c}, we plot the energies of the charged and uncharged states
for voltage parameters that place the molecule closely below the Fermi level of the source
electrode, that is for the same $V_\text{g}$ as before but with opposite sign. In both processes,
tunneling onto and off the molecule, there is no state of the charged system that is isolated due to
the selection rule or the spectral distortion. The property of the Jahn-Teller effect to
\textit{lower} the energy of the distorted states is the actual reason for this asymmetry.  If our
Jahn-Teller molecule were such that it was Jahn-Teller active \textit{without} an excess charge and
symmetric \textit{with} an additional electron, the situation would be inverted, current-blockade for
positive, conducting at negative gate voltage.

\section{Rate equations}\label{sec:Rateequations}

In the previous paragraphs, we have argued that due to the interplay of the spectral distortion and
the selection rule, there are two time scales present in the problem. One describes the tunnel
dynamics and is of order $\mathcal{O}(H_\text{T}^2)$. The other is due to higher-order processes and
is an estimate of how long the system spends in the trapping states. It is thus evident that the
precise implementation of the internal dynamics is irrelevant as long as the key ingredients to the
separation of time scales, namely selection rule and spectral distortion, are included.

In this paragraph, we present such a numerical implementation that accounts for the selection rule by
employing rate equations in the pseudo angular momentum basis. By doing so, we neglect all
coherences between \textit{any} two states in the density matrix, and therefore also those between
degenerate states. This will, as we have pointed out before, \textit{not} affect the appearance and
structure of the current-blockade, because the effect is independent of the specific form of the
eigenbasis of the stationary density matrix. The only error that is introduced by the restriction to
rate equations is a quantitative one. The rate equations, on the contrary, allow for a very intuitive
and generic treatment of transport problems for systems with spectral distortions and selection
rules.

\subsection{Formalism}
The occupation probability $P^n_q$ of the molecular eigenstates $\ket{n;q}$ ($n=0,1$ denotes
the charge of the molecule, and $q$ is a multiindex for the internal degrees of freedom. $\Gamma$ is
the broadening of the molecular levels due to the coupling to the leads.) is governed by the rate
equation\cite{Mitra04} 
\[\label{eq:RATE} 
\dot{P}^n_q = \sum_{n', q'} \left[P^{n'}_{q'}W^{n'\rightarrow n}_{q'\rightarrow q} - P^n_q
W^{n\rightarrow n'}_{q\rightarrow q'}\right] + \gamma_{\rm rel}.
\]
The tunnel rates $W^{n\rightarrow n'}_{q\rightarrow q'}$ are a product of the Golden-rule
expressions Eq.~(\ref{eq:GoldenRule}) and the Fermi functions of the leads.
\cite{Mitra04,Koch04} The last term, $\gamma_{\rm rel}= - ({1}/{\tau})[P^n_q - P^{\text{eq}}_q
\sum_{q'}P^n_{q'}]$, describes vibrational relaxation towards the thermal equilibrium distribution
$P^{\text{eq}}_q$ by higher order processes not included in the Hamiltonian. By doing so, we effectively
implement the aforementioned processes, which release the system from the trapping states. 
Our results, however, are independent of the details of this implementation. The maximal
tunneling rate, $\tau_0^{-1}$, defines the proper time scale on which we measure all times.

The stationary solution of eq.~(\ref{eq:RATE}) permits us to compute the stationary current through,
say, the left junction,\footnote{The index L on the rates indicate that only this lead's Fermi
function is used for the computation of the rate.} 
\[
I_\text{L} = \sum_{qq'}\bigl[ P^0_q W^{0\rightarrow 1}_{q\rightarrow q';\text{L}} - P^1_q
W^{1\rightarrow 0}_{q\rightarrow q';\text{L}}\bigr],
\]
as a function of  the source-drain voltage $V_\text{sd}$ and the gate voltage $V_\text{g}$. The
results for the differential conductance $dI/dV_\text{sd}$ shown in Fig.\ \ref{fig:results} for (a)
fast and (b) slow vibrational relaxation differ strongly from those obtained for single-mode
models.~\cite{Mitra04} When the molecular orbitals are close to the Fermi energy of the drain
electrode (negative $V_\text{g}$ in Fig.\ \ref{fig:results}), we find, in accordance with the
discussion in section \ref{sec:Theory}, an extended region of current suppression outside the
Coulomb-blockaded region with large negative differential conductance (NDC). No such suppression
occurs when the molecular orbitals are close to the chemical potential of the source electrode
(positive $V_\text{g}$ in Fig.~\ref{fig:results}), thus rendering $dI/dV$ asymmetric in the gate
voltage. The positions of the $dI/dV$-lines that mark the area of the suppressed current in the
diagrams are defined by those values of $(V_\text{sd}, V_\text{g})$ for which the trapping state is
connected to an energy manifold that participates in the transport process. If an increase of the
bias voltage generates a finite escape rate for population from the trapping state, that state is no
longer absorbing, and a stationary current can flow, hence a line of \textit{positive} differential
conductance is observed. In the opposite case, that is trapping behaviour, a line of
\textit{negative} differential conductance appears as soon as the principal current carrying
states---most often it is sufficient to consider the vibronic ground state of the system---have
tunnel rates into the trapping state. Because trapping states are excited vibronic states and
intermediate excited states are necessary to access these, there is a certain energy threshold below
which a finite current may flow but the trapping state is still out of reach. This threshold
will thus be visible as a negative differential conductance peaks in the $dI/dV$ diagram. Since the trapping states with
larger pseudo angular momentum have larger energy thresholds before they can be populated in steady
state, the current-blockade mechanism will yield a step-like structure of negative differential
conductance peaks and plateaus of suppressed current. These are clearly visible in our numerical
results in Fig.~\ref{fig:results}.  

\subsection{Separation of time scales}

We further support the irrelevance of the details of the fast, order $\mathcal{O}(H_\text{T}^2)$
dynamics by reproducing the positions and line shapes of the peaks in the numerically obtained
differential conductance, see Fig.~\ref{fig:results}, within a simple model. This model only retains the most basic features of the dynamics, namely the
presence of trapping states and the two relevant time scales. Consider a two-level
system with a trapping state $\ket{\beta}$ and a state $\ket{\alpha}$ from which the electron can
tunnel to the drain electrode, say the absolute ground state of the charged molecule. The current is
given by the inverse of the time $\tau_\text{total}$ the electron spends on the molecule. In our model,
this time is the sum of the trapping time $\tau$ and the time the electron needs to leave the
second, non-trapping state. The rate to leave this latter state is the product of the bare tunneling
rate $1/\tau_0$ and the probability $1-f$ to find an empty state of appropriate energy in the drain
electrode, where $f$ is this electrode's Fermi function. The typical time for tunneling from
$\ket{\alpha}$ out to the leads is thus $\tau_0/(1-f)$ and the stationary current can be estimated
as
\[\label{eq:I_model}
I \propto \frac{1}{\tau_\text{total}} = \frac{1}{\tau + \frac{\tau_0}{(1-f)}} = \frac{1-f}{(1-f)\tau + \tau_0}.
\]
This yields the relation $dI/dV \propto I (\tau I - 1)f/T$. Its consequences for the
line-shape are twofold. As a purely quantitative effect, the height of the peak in $dI/dV$ decreases
with increasing $\tau$. As a qualitative effect, however, the position of the peak is shifted. In a model with
just a single time scale, be it a quantum dot or a single-mode molecule, the stationary current
through the drain electrode is \textit{only} determined by the Fermi function $1-f(\varepsilon_{1;q}
- \varepsilon_{0;q'} -\mu_\alpha)$, that is the probability to find an empty state in the leads with
just the energy necessary for the vibrational transition $\ket{0;q'} \leftrightarrow \ket{1;q}$. The
maximum of the $dI/dV$-peak is thus located at the bias voltage for which $\mu_\alpha =
\varepsilon_{1;q} - \varepsilon_{0;q'}$. Within our two-state model, this behavior is recovered for
$\tau=0$ which corresponds to the absence of any trapping effect due to fast relaxation. When trapping is
included, and hence $\tau$ is finite, the maximum of the differential conductance peak is shifted.

\begin{figure}
\begin{minipage}{.8\linewidth}
    \includegraphics[width=\linewidth]{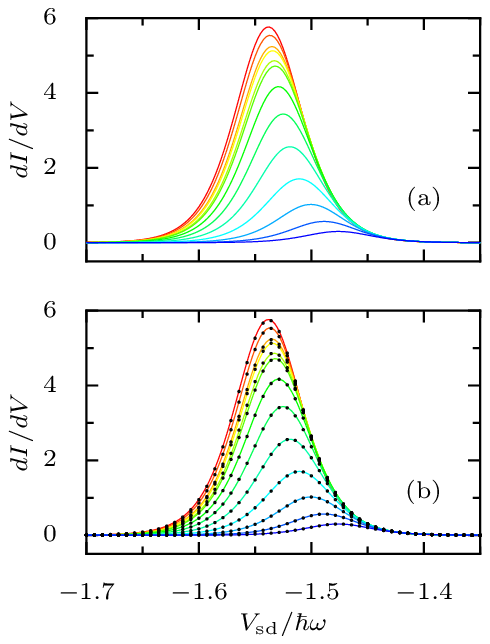}
 \end{minipage}
 \caption{Positive differential conduction peak at the Coulomb-blockade, $V_\text{g} =
 -0.5\hbar\omega$, for different values of the relaxation time $\tau$ (a) and fit with the
 simplified model, dots in (b). The read curves, that is the large, centered peaks have small, the
 blue ones, damped and shifted to lower $V_\text{sd}$ very large $\tau$.\label{fig:Relaxation}}
\end{figure}

\begin{figure}[b]
\begin{minipage}{.8\linewidth}
    \includegraphics[width=\linewidth]{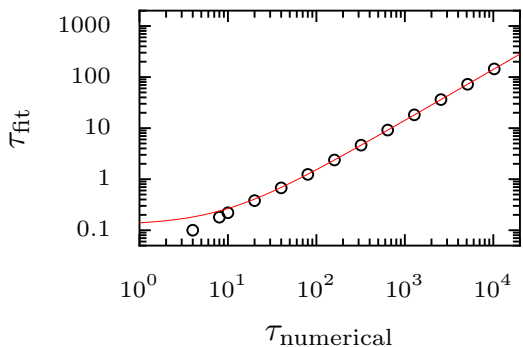}
 \end{minipage}
 \caption{Fit $\tau$-values of the simplified model (dots) and its application to Fig.~\ref{fig:Relaxation}
 vs. the numerical values used in the actual simulation. The curve is a linear fit to these
 values, $\tau_\text{fit} \propto \tau_\text{numerical}$,  showing the validity of the simplified approach in
 the regime of strong separation of the time scales.\label{fig:Relaxation_fit}}
\end{figure}

This simple model compares remarkably well to the numerical rate-equation results of the previous
section.  We focus on a single peak in the $dI/dV$ diagram and study its line shape for different
values of the phenomenological relaxation rate $\tau$ of Eq.~(\ref{eq:RATE}). The effect of
increasing $\tau$---red to blue curves in Fig.~\ref{fig:Relaxation}~(a)---is a reduction of the
peak's height and a shift of its position. We can fit the results of our simple model,
Eq.~(\ref{eq:I_model}), to the data in Fig.~\ref{fig:Relaxation} (a), obtaining
Fig.~\ref{fig:Relaxation} (b). Comparing the fit values for $\tau$ with the parameters of the
numerical simulation, see Fig.~\ref{fig:Relaxation_fit}, we can see how well this simplified model
describes the interplay of the two time scales. The only deviations occur when both time scales are
of comparable magnitude, $\tau = \mathcal{O}(\tau_0)$. In contrast, for a sufficiently large
separation of time scales, the agreement is nearly perfect. This backs our general considerations of
section \ref{sec:Theory}, where we have argued that the observed phenomena are of a generic nature.


The predominant features of electronic transport through a single Jahn-Teller molecule are
determined by three properties of the system: the spectral distortion and the selection rule fix the
position of the peaks in the differential conductivity, their heights are determined by the
Franck-Condon matrix elements. Finally, higher order processes introduce a second, larger time scale, which
eventually determines the exact line shape. 

\section{Experimental issues}\label{sec:Perturbations}

We finally discuss several experimental issues. Placing a Jahn-Teller molecule between two biased
electrodes exposes it to a potentially large electric field. For a semiconductor quantum dot, weak
coupling and good screening implies that the potential drop is concentrated in the contact regions
between the electrodes and the dot.\cite{Ratner03} For a single molecule, however, this is less obvious. Thus, the
electric field may lift the electronic degeneracy by the Stark effect and mix the vibronic levels by
breaking the rotational symmetry of the adiabatic potential. 

To estimate the mixing of vibronic levels, we note that the relevant vibrational modes $\theta$ and
$\varepsilon$ do not possess a dipole moment and are therefore unaffected by the electric field
unless anharmonic mode mixing becomes relevant when other, ``dipole-active'' modes are strongly
displaced. One readily estimates, however, that for relevant electric fields (corresponding to
voltage drops of the order of $\hbar\omega$) the typical displacements of dipole-active vibrational
modes are small compared with their oscillator length $\ell_{\rm osc}$, so that no anharmonic
effects are expected. Assuming that the dipole moment is of order $ex$, where $x$ is the
displacement of the dipole-active mode, we obtain a displacement $x\sim eE/M\omega^2\sim
\ell^2_\text{osc}/\delta \ll \ell_\text{osc}$. The length $\delta$ is of the order of the size of
the molecular junction. We also note that the electric field does not directly couple the two
electronic levels as both are $d$ orbitals. The Stark shift is therefore quadratic in the electric
field, and we estimate that it is several orders of magnitude smaller than the vibrational
frequency.

The coupling of the Jahn-Teller molecule to fermionic reservoirs, the electronic leads, may
also introduce external perturbations that are able to break the symmetries of the system on
which our approach relies so heavily. If the perturbation is too strong, for example, the degeneracy
of the vibrational modes $\theta$ and $\varepsilon$ is lifted and the energies are shifted by a
large amount. The Jahn-Teller effect is then no longer of $\Exe$- but rather of $E\!\otimes\!(b_1 +
b_2)$-type. A selection rule like the one used in this paper is not available in such a model, and
the current-blockade would not be visible. If, however, the perturbation is only small compared with
the tunnel coupling, the mixing of the states induced by the perturbation is able to introduce a
second time scale $\tau\gg\tau_0$ into the problem. As we have discussed in the previous paragraphs,
the presence of two widely separated time scales only shifts the positions and alters the shape of
the differential conductance peaks, no matter what the detailed physics behind the phenomenology of
the conduction peaks are.

\section{Conclusions}\label{sec:Conclusion} 

We have shown that symmetry-induced degeneracies of electronic and vibrational modes can lead to new
transport phenomena in single-molecule junctions. When the molecular junction consists of, say, an
octahedral complex with an $\Exe$ Jahn-Teller effect, a combination of Berry-phase effects and a
spectral trapping mechanism results in a non-trivial current-blockade, even for parameters where the
Coulomb blockade is lifted. It is evident from the physics of this blockade mechanism
developed in this paper that it is associated with strongly enhanced shot noise of the
current.~\cite{unpublished} We have also shown that the mechanism is quite generic and that the
introduction of a second time scale to the problem, for example by higher order relaxation
processes, co-tunneling, or external perturbations, has only quantitative effects on our results.

The good agreement of the numerical results of our effective rate model with the general
consideration of section \ref{sec:Theory} shows that the current-blockade that has been found in
this context is a generic effect of systems with selection rules and spectral distortions.  Nowack
\textit{et al}.\cite{Nowack05} find a related current-blockade effect in systems where the strong
spectral distortion alone damps certain Franck-Condon matrix elements. This is, however, no
selection rule that derives from symmetries like in this paper; it rather stems from the almost vanishing overlap of
displaced oscillator wave functions.

Our discussion of Jahn-Teller molecules raises several interesting questions. The most important
question concerns a complete understanding of the role of coherences between degenerate states and
their implications for the transport properties of the system. What are the experimentally
measurable quantities that are influenced by non-diagonal elements of the density matrix? And when
is a reduction to rate equations allowed? 

\section*{Acknowledgments}
This work was supported in part by SPP 1243 of the
Deutsche Forschungsgemeinschaft as well as Sfb 658.


\begin{thebibliography}{10}

\bibitem{Park00} H.\ Park, J.~Park, A.~K.~L.~Lim, E.~H.~Anderson, A.~P.~Alvisatos, and P.~L.~McEuen, Nature {\bf 407}, 57 (2000).

\bibitem{Ruitenbeek_H2} R.~H.~M.\ Smit, Y.~Noat, C.~Untiedt, N.~D.~Land, M.C~van~Hemert, and
  J.~M.~van Ruitenbeek, Nature {\bf 419}, 906 (2002).

\bibitem{Ralph} J.~Park, A.~N.~Pasupathy, J.~I.~Goldsmith, C.~Chang, Y.~Yaish, J.~R.~Petta,
  M.~Rinkoski, J.~P.~Sethna, H.~D.~Abru\~na, P.~L.~McEuen, and D.~C.~Ralph, Nature {\bf 417}, 722 (2002).

\bibitem{Zant} H.~B.\ Heersche, Z.~de~Groot, J.~A.~Folk, H.~S.J~van~der~Zant, C.~Romeike,
  M.~R.~Wegewijs, L.~Zobbi, D.~Barreca, E.~Tondello, and A.~Cornia, Phys.\ Rev.\ Lett.\ {\bf 96}, 206801 (2006).

\bibitem{Yacoby} T.~Dadosh, Y.~Gordin, R.~Krahne, I.~Khivrich, D.~Mahalu, V.~Freydman, J.~Sperling,
  A.~Yacoby, and I.~Bar-Joseph, Nature {\bf 436}, 677 (2005).

\bibitem{Natelson} L.~H.~Yu, Z.~K.~Keane, J.~W.~Ciszek, L.~Cheng, J.~M.~Tour, T.~Baruah,
  M.~R.~Pederson, and D.~Natelson, Phys.\ Rev.\ Lett.\ {\bf 95}, 256803 (2005).

\bibitem{Braig03} S.\ Braig and K.\ Flensberg, Phys.\ Rev.\ B {\bf 68},
  205324 (2003).

\bibitem{Mitra04} A.~Mitra, I.~Aleiner, and A.~J.~Millis, Phys.~Rev.~B
  {\bf 69}, 245302 (2004).

\bibitem{Gorelik98} L.~Y.~Gorelik and {\it et al.},
  Phys.\ Rev.\ Lett.\ {\bf 80}, 4526 (1998).

\bibitem{Koch05} J.~Koch and F.~von~Oppen, Phys.~Rev.~Lett.~{\bf 94},
  206804 (2005).

\bibitem{Koch06} J.\ Koch, M.E.\ Raikh, and F.\ von Oppen,
  Phys.\ Rev.\ Lett.\ {\bf 96}, 056803 (2006).

\bibitem{Donarini06} A.\ Donarini, M.\ Grifoni, and K.\ Richter,
  Phys.\ Rev.\ Lett.\ {\bf 97}, 166801 (2006).

\bibitem{Nowack05} K.~C.\ Nowack and M.R.\ Wegewijs, cond-mat/0506552.

\bibitem{Kaat04} G.A.\ Kaat and K.\ Flensberg,
  Phys.\ Rev.\ B\ {\bf 71}, 155408 (2005).

\bibitem{Nitzan} For a recent review, see M.\ Galperin, M.~A.\ Ratner, and A.\ Nitzan, cond-mat/0612085.

\bibitem{Blum81} K.~Blum, \textit{Density Matrix Theory and Applications}, Plenum Press, New York
  and London (1981).

\bibitem{Bersuker} I.~B.\ Bersuker, \textit{The Jahn-Teller Effect},
  Cambridge University Press, Cambridge (2006).

\bibitem{Berry84} M.~V.\ Berry, Proc.\ R.\ Soc.\ Lond.\ Ser.\ A, {\bf 392},
45 (1984).

\bibitem{Longuet} H.~.C.~Longuet-Higgins, U.~{\"O}pik, M.~H.~L.~Pryce, and R.~A.~Sachs,
  Proc.~R.~Soc.~London.~A, {\bf 244}, 1--15 (1958)

\bibitem{Koch04} J.~Koch, F.~von~Oppen, Y.~Oreg,~and E.~Sela, Phys.~Rev.~B~{\bf 70}, 195107 (2004)

\bibitem{Ratner03} Y.~Xue and M.~A.~Ratner, Phys.~Rev.~B~{\bf 68}, 115406 (2003)

\bibitem{unpublished} M.~G.\ Schultz, T.~S.\ Nunner, and F.\ von Oppen, unpublished.

\end{thebibliography}
\end{document}